%% file: asilomar22.tex
\documentclass[conference]{IEEEtran}
\IEEEoverridecommandlockouts

\usepackage{cite}
\usepackage{amsmath,amssymb,amsfonts}
\usepackage{algorithmic}
\usepackage{algorithm}
\usepackage{graphicx}
\usepackage{textcomp}
\usepackage{xcolor}
\usepackage{siunitx}
\usepackage{tikz}
\usepackage{pgfplots}
\usepgfplotslibrary{statistics}
\pgfplotsset{compat=1.8}
\usepackage{caption}
\usepackage{balance}

\newcommand{\smallplotheight}{0.575\columnwidth} 
\newcommand{\smallplotwidth}{0.5\columnwidth}
\newcommand{\normalplotheight}{0.4\columnwidth}
\newcommand{\normalplotwidth}{1.0\columnwidth}

\newcommand{\lineWidth}{1.0pt}
\newcommand{\markSize}{2.0pt}
\definecolor{ourdarkblue}{RGB}{30, 100, 200}
\definecolor{ourdarkgreen}{RGB}{0, 100, 0}
\definecolor{ouryellow}{RGB}{220, 210, 50}

\tikzset{unipowcdf/.style={mark options={solid}, color=ourdarkgreen, line width=\lineWidth, mark=None, mark size=\markSize, dotted}}
\tikzset{upchansubcdf/.style={mark options={solid}, color=purple, line width=\lineWidth, mark=None, mark size=\markSize, dotted}}

\tikzset{lloydpgddl/.style={mark options={solid}, color=orange, line width=\lineWidth, mark=None, mark size=\markSize}}
\tikzset{lloydpgdul/.style={mark options={solid}, color=cyan, line width=\lineWidth, mark=None, mark size=\markSize, dotted}}
\tikzset{lloydlaudl/.style={mark options={solid}, color=red, line width=\lineWidth, mark=None, mark size=\markSize}}
\tikzset{lloydlauul/.style={mark options={solid}, color=green, line width=\lineWidth, mark=None, mark size=\markSize, dotted}}
\tikzset{gmmpgddl/.style={mark options={solid}, color=blue, line width=\lineWidth, mark=None, mark size=\markSize}}
\tikzset{gmmpgdul/.style={mark options={solid}, color=brown, line width=\lineWidth, mark=None, mark size=\markSize, dotted}}
\tikzset{gmmlaudl/.style={mark options={solid}, color=gray, line width=\lineWidth, mark=None, mark size=\markSize}}
\tikzset{gmmlauul/.style={mark options={solid}, color=black, line width=\lineWidth, mark=None, mark size=\markSize, dotted}}

\newcommand{\pltHeigsp}{uni pow eigsp}
\newcommand{\pltUniPowCov}{uni pow cov}
\newcommand{\pltlloydpgddl}{Lloyd PGD DL}
\newcommand{\pltlloydpgdul}{Lloyd PGD UL}
\newcommand{\pltlloydlaudl}{Lloyd Lau DL}
\newcommand{\pltlloydlauul}{Lloyd Lau UL}
\newcommand{\pltgmmpgddl}{GMM PGD DL}
\newcommand{\pltgmmpgdul}{GMM PGD UL}
\newcommand{\pltgmmlaudl}{GMM Lau DL}
\newcommand{\pltgmmlauul}{GMM Lau UL}

\tikzset{lloydpgdulperfect/.style={mark options={solid}, color=black, line width=\lineWidth, mark=None, mark size=\markSize, dashed}}
\tikzset{lloydpgdulomp/.style={mark options={solid}, color=orange, line width=\lineWidth, mark=None, mark size=\markSize, dashed}}
\tikzset{lloydpgdulls/.style={mark options={solid}, color=red, line width=\lineWidth, mark=None, mark size=\markSize}}
\tikzset{lloydpgdulscov/.style={mark options={solid}, color=red, line width=\lineWidth, mark=None, mark size=\markSize, dashed}}
\tikzset{lloydpgdulgmmulbest/.style={mark options={solid}, color=gray, line width=\lineWidth, mark=None, mark size=\markSize}}
\tikzset{lloydpgdulgmmulall/.style={mark options={solid}, color=gray, line width=\lineWidth, mark=None, mark size=\markSize, dashed}}

\tikzset{gmmpgdulperfect/.style={mark options={solid}, color=blue, line width=\lineWidth, mark=None, mark size=\markSize}}
\tikzset{gmmpgdulfromy/.style={mark options={solid}, color=green, line width=\lineWidth, mark=None, mark size=\markSize}}

\newcommand{\pltlloydpgdulperfect}{Lloyd PGD, $\mbh$}
\newcommand{\pltlloydpgdulomp}{Lloyd PGD, $\hhat_{\text{OMP}}$}

\newcommand{\pltlloydpgdulscov}{Lloyd PGD, $\hhat_{\text{s-cov}}$}

\newcommand{\pltlloydpgdulgmmulall}{Lloyd PGD, $\hhat_{\text{GMM}}^{(K)}$}
\newcommand{\pltgmmpgdulperfect}{GMM PGD, $\mbh$}
\newcommand{\pltgmmpgdulfromy}{GMM PGD, $\mby$}

\def\BibTeX{{\rm B\kern-.05em{\sc i\kern-.025em b}\kern-.08em
    T\kern-.1667em\lower.7ex\hbox{E}\kern-.125emX}}

\input{miko}

\Crefname{figure}{Fig.}{Figs.}

\newacronym{AWGN}{AWGN}{additive white Gaussian noise}
\newacronym{BLMMSE}{BLMMSE}{Bussgang LMMSE}
\newacronym{BS}{BS}{base station}
\newacronym{CDF}{CDF}{cumulative distribution function}
\newacronym{CNN}{CNN}{convolutional neural network}
\newacronym{CSI}{CSI}{channel state information}
\newacronym{CSIT}{CSIT}{channel state information at the transmitter}
\newacronym{DFT}{DFT}{discrete Fourier transform}
\newacronym{DL}{DL}{downlink}
\newacronym{DNN}{DNN}{deep neural network}
\newacronym{DoA}{DoA}{direction of arrival}
\newacronym{EM}{EM}{expectation maximization}
\newacronym{FDD}{FDD}{frequency division duplex}
\newacronym{GMM}{GMM}{Gaussian mixture model}
\newacronym{LMMSE}{LMMSE}{linear minimum mean square error}
\newacronym{LOS}{LOS}{line of sight}
\newacronym{LS}{LS}{least squares}
\newacronym{MSE}{MSE}{mean squared error}
\newacronym{MIMO}{MIMO}{multiple-input multiple-output}
\newacronym{MPC}{MPC}{multi-path component}
\newacronym{MT}{MT}{mobile terminal}
\newacronym{NLOS}{NLOS}{non-line of sight}
\newacronym{NN}{NN}{neural network}
\newacronym{O2I}{O2I}{outdoor-to-indoor}
\newacronym{OMP}{OMP}{orthogonal matching pursuit}
\newacronym{PDF}{PDF}{probability density function}
\newacronym{PGD}{PGD}{projected gradient descent}
\newacronym{PSD}{PSD}{power spectral density}
\newacronym{SNR}{SNR}{signal-to-noise ratio}
\newacronym{TDD}{TDD}{time division duplex}
\newacronym{UL}{UL}{uplink}
\newacronym{ULA}{ULA}{uniform linear array}
\newacronym{URA}{URA}{uniform rectangular array}
\newacronym{UMa}{UMa}{urban macrocell}
\newacronym{nSE}{nSE}{normalized spectral efficiency}
\newacronym{cCDF}{cCDF}{complementary cumulative distribution function}

\pgfplotsset{compat=1.15}

\newcommand{\Ncbentries}{K}
\newcommand{\Nrx}{N_{\mathrm{rx}}}
\newcommand{\Ntx}{N_{\mathrm{tx}}}
\newcommand{\Ntxv}{N_{\mathrm{tx,v}}}
\newcommand{\Ntxh}{N_{\mathrm{tx,h}}}
\newcommand{\Krx}{K_{\mathrm{rx}}}
\newcommand{\Ktx}{K_{\mathrm{tx}}}
\newcommand{\Ttr}{T_{\mathrm{train}}}

\begin{document}




\title{GMM-based Codebook Construction and \\ Feedback Encoding in FDD Systems\thanks{\copyright This work has been submitted to the IEEE for possible publication. Copyright may be transferred without notice, after which this version may no longer be accessible.
}}

\author{\centerline{Nurettin Turan${^*}$, Michael Koller$^*$, Benedikt Fesl$^*$, Samer Bazzi$^\dagger$, Wen Xu$^\dagger$, and Wolfgang Utschick$^*$ }\\
\IEEEauthorblockA{$^*$Professur f\"ur Methoden der Signalverarbeitung, Technische Universit\"at M\"unchen, 80333 Munich, Germany\\ 
$^\dagger$Huawei Technologies Duesseldorf GmbH, 80992 Munich, Germany\\
Email: \small{{\{nurettin.turan,michael.koller,benedikt.fesl,utschick\}@tum.de}}\\Email: \small{{\{samer.bazzi,wen.dr.xu\}@huawei.com}}}
}

\maketitle

\begin{abstract}
We propose a precoder codebook construction and feedback encoding scheme which is based on \acp{GMM}.
In an offline phase, the \ac{BS} first fits a \ac{GMM} to \ac{UL} training samples.
Thereafter, it designs a codebook in an unsupervised manner by exploiting the \ac{GMM}'s clustering capability.
We design one codebook entry per \ac{GMM} component.
After offloading the \ac{GMM}---but not the codebook---to the \ac{MT} in the online phase, the \ac{MT} utilizes the \ac{GMM} to determine the best fitting codebook entry.
To this end, no channel estimation is necessary at the \ac{MT}. Instead, the \ac{MT}'s observed signal is used to evaluate how responsible each component of the \ac{GMM} is for the signal. The feedback consists of the index of the \ac{GMM} component with the highest responsibility and the \ac{BS} then employs the corresponding codebook entry.
Simulation results show that the proposed codebook design and feedback encoding scheme outperforms conventional Lloyd clustering based codebook design algorithms, especially in configurations with reduced pilot overhead.
\end{abstract}

\begin{IEEEkeywords}
Gaussian mixture models, machine learning, feedback, codebook design, frequency division duplexing
\end{IEEEkeywords}

\section{Introduction}

In \ac{MIMO} communications systems, \ac{CSI} has to be acquired at the \ac{BS} in regular time intervals.
In \ac{FDD} mode, the \ac{BS} and the \ac{MT} transmit in the same time slot but at different frequencies. 
This breaks the reciprocity between the instantaneous \ac{UL} \ac{CSI} and \ac{DL} \ac{CSI}.
Accordingly, acquiring \ac{DL} \ac{CSI} in \ac{FDD} operation is difficult \cite{2019massive}.
Possible solutions include to either extrapolate the \ac{DL} \ac{CSI} from the estimate of the \ac{UL} \ac{CSI} at the BS, or to transfer the \ac{DL} \ac{CSI} estimated at the \ac{MT} to the \ac{BS} directly or in a highly compressed version.
However, the most common solution in practice is to avoid the direct feedback of the \ac{CSI} and to use only a small number of feedback bits. For instance, as it is done in this paper, the feedback can be used as an index into a predefined codebook of precoders \cite{Love}.

In recent years, machine learning techniques have been explored in the context of communications.
However, these typically need a large dataset of \ac{DL} channels for their training phases.
This would require the users to collect large amounts of \ac{DL} \ac{CSI} and either to perform the training themselves or to share the collected data with the \ac{BS}.
The corresponding computation and/or signaling overhead involved in such a scheme is generally unaffordable in practice.

Recently, in \cite{utschick2021} it has been shown that \ac{DL} \ac{CSI} training data can be replaced with \ac{UL} \ac{CSI} training data even for the design of \ac{DL} functionalities.
This completely eliminates the aforementioned overhead.
\ac{UL} \ac{CSI} can, e.g., be acquired at the \ac{BS} during the regular \ac{UL} transmission.
In \cite{utschick2021}, the observation has been made in the context of training autoencoders. Similar observations have since been made in~\cite{fesl2021centralized} for \ac{DL} channel estimation and in~\cite{TuKoBaXuUt21, TuKoRiFeBaXuUt21} for codebook design.
In this work, we also utilize the idea to centrally learn \ac{DL}-related functionalities at the \ac{BS} using \ac{UL} training data.

The contributions of this work are summarized as follows.
We propose a codebook construction and feedback encoding scheme which is based on \acp{GMM}.
Since \acp{GMM} are universal approximators~\cite{NgNgChMc20}, we can use a \( K \)-components \ac{GMM} to approximate the unknown channel \ac{PDF}.
In the \emph{offline} phase, we propose to fit the \ac{GMM} centrally at the \ac{BS} using solely \ac{UL} data. Thereafter, we cluster the training data according to the \ac{GMM} components and design a codebook entry for every component.
In this way, the codebook is designed for a whole scenario, i.e., for the whole site in which the \ac{BS} is located.
The \ac{GMM} is offloaded to every \ac{MT} in the coverage area of the \ac{BS}, with which the \ac{MT} can then select the best fitting codebook entry in the \emph{online} phase.

In conventional approaches, first the \ac{DL} \ac{CSI} is estimated and subsequently the best fitting codebook entry is determined.
The technique proposed in this paper allows to bypass explicit \ac{CSI} estimation and yields the best fitting codebook entry by using the receive signal to evaluate the \ac{GMM} responsibilities. 
The \( k \)th responsibility of a \ac{GMM} corresponds to the probability that the \( k \)th \ac{GMM} component is responsible for the receive signal.
The feedback then consists of the index \( k \) of the component with the highest responsibility.
Finally, we make use of a Kronecker approximation to significantly reduce the number of \ac{GMM} parameters
such that the offloading overhead is smaller.
In simulations, the proposed codebook design and feedback encoding scheme outperforms conventional Lloyd clustering based codebook design algorithms \cite{LiBuGr80, LaYoCh04}.

\section{System Model}

The \ac{DL} received signal of a point-to-point \ac{MIMO} system can be expressed as $\mby^\prime = \mbH \mbx + \mbn^\prime$,
where $\mby^\prime \in \C^{\Nrx}$ is the receive vector, $\mbx \in \C^{\Ntx}$ is the transmit vector sent over the \ac{MIMO} channel $\mbH \in \C^{\Nrx \times \Ntx}$, and $\mbn^\prime \sim \mathcal{N}_\C(\mathbf{0},
\sigma_n^2 \mbI_{\Nrx})$ denotes the \ac{AWGN}.
In this paper, we consider system configurations with $\Nrx < \Ntx$. The \ac{BS} is equipped with a \ac{URA} and the \ac{MT} is equipped with a \ac{ULA}.
If perfect \ac{CSI} is known to both the transmitter and receiver, and assuming input data with Gaussian distribution, the capacity of the \ac{MIMO} channel is \cite{Goldsmith, Goldsmith2}:
\begin{equation}
    C = \max_{\mbQ \succeq \mbzero, \tr\mbQ\leq\rho} \log_2 \det\left( \mbI + \frac{1}{\sigma_n^2} \mbH \mbQ \mbH^\herm\right),
\end{equation}
where $\mbQ \in \C^{\Ntx \times \Ntx}$ is the transmit covariance matrix and the transmit vector is then given by \( \mbx = \mbQ^{1/2} \mbs \) with \( \expec[\mbs \mbs^\herm] = \mbI_{\Ntx} \)~\cite{Love}. 
The optimal transmit covariance matrix $\mbQ^\star$ of the link between the \ac{BS} and a \ac{MT} achieves the capacity and can be obtained by decomposing the channel into $\Nrx$ parallel streams and employing water-filling \cite{Telatar99capacityof}. 

Channel reciprocity can generally not be assumed in \ac{FDD} systems, e.g., \cite{Love}.
Therefore, only the \ac{MT} could compute the optimal transmit covariance matrix \( \mbQ^\star \) if it estimated the \ac{DL} \ac{CSI}. 
This makes some form of feedback from the \ac{MT} to the \ac{BS} necessary.
Ideally, the user would feed the complete \ac{DL} \ac{CSI} back to the \ac{BS}, which is infeasible in general.
Instead, typically limited feedback is considered where a small number of \( B \) bits is sent back to the \ac{BS}.
The \( B \) feedback bits are typically used for encoding an index that specifies an element from a set of covariance matrices.
That is, the \ac{MT} and \ac{BS} share a {codebook} $ \mathcal{Q} = \{\mbQ_1, \mbQ_2, \dots, \mbQ_{2^B} \} $ of $ 2^B $ pre-computed transmit covariance matrices, and the \ac{MT} is assumed to estimate the \ac{DL} channel \( \mbH \) and then uses it to determine the best codebook entry \( \mbQ_{k^\star} \) via
\begin{equation}\label{eq:codebook_index_selection}
    k^\star = \argmax_{k \in \{1, \dots, 2^B\}} \log_2 \det\left( \mbI + \frac{1}{\sigma_n^2} \mbH \mbQ_k \mbH^\herm\right).
\end{equation}
The feedback consists of the index \( k^\star \) encoded by \( B \) bits, and the \ac{BS} employs the transmit covariance matrix \( \mbQ_{k^\star} \) for data transmission.

\section{Channel Model and Data Generation}

Version $2.6.1$ of the QuaDRiGa channel simulator \cite{QuaDRiGa1, QuaDRiGa2} is used to generate \ac{CSI} for the \ac{UL} and \ac{DL} domains in an \ac{UMa} scenario.
The carrier frequencies are $\SI{2.53}{\giga\hertz}$ for the \ac{UL} and $\SI{2.73}{\giga\hertz}$ for the \ac{DL} such that there is a frequency gap of $\SI{200}{\mega\hertz}$.
The \ac{BS} uses a \ac{URA} with ``3GPP-3D'' antennas, and the \acp{MT} use \acp{ULA} with ``omni-directional'' antennas.
The \ac{BS} covers a $\SI{120}{\degree}$ sector and is placed at $\SI{25}{\meter}$ height.
The minimum and maximum distances between \acp{MT} and the \ac{BS} are $\SI{35}{\meter}$ and $\SI{500}{\meter}$, respectively.
In $80\%$ of the cases, the \acp{MT} are located indoors at different floor levels.
The outdoor \acp{MT} have a height of $\SI{1.5}{\meter}$.%

Many parameters such as path-loss, delay, and angular spreads, path-powers for each subpath, and antenna patterns are different in the \ac{DL} and \ac{UL} domain~\cite{QuaDRiGa1}.
However, the following parameters are identical:
\ac{BS} location and the \ac{MT} locations, propagation cluster delays and angles for each \ac{MPC}, and the spatial consistency of the large scale fading parameters.
A QuaDRiGa \ac{MIMO} channel is given by $\mbH = \sum_{\ell=1}^{L} \mbG_{\ell} e^{-2\pi j f_c \tau_{\ell}}$
with $\ell$ the path number, $L$ the number of \acp{MPC}, $f_c$ the carrier frequency, and $\tau_{\ell}$ the \( \ell \)th path delay.
The number \( L \) depends on whether there is \ac{LOS}, \ac{NLOS}, or \ac{O2I} propagation: $L_\text{LOS} = 37$, $L_\text{NLOS} = 61$ or $L_\text{O2I} = 37$.
The coefficients matrix $\mbG_{\ell}$ consists of one complex entry for each antenna pair and comprises the attenuation of a path, the antenna radiation pattern weighting, and the polarization.
As described in the QuaDRiGa manual~\cite{QuaDRiGa2}, the generated channels are post-processed to remove the path gain.

\section{Channel Estimation}


In the pilot transmission phase, the \ac{DL} received signal is:
\begin{equation} \label{eq:noisy_obs}
    \mbY = \mbH \mbP + \mbN \in \C^{\Nrx \times n_p} ,
\end{equation}
where $n_p$ is the number of transmitted pilots and $\mbN = [\mbn^{\prime}_1, \mbn^{\prime}_2, \dots, \mbn^{\prime}_{n_p}] \in \C^{\Nrx \times n_p}$. The pilot matrix $\mbP \in \C^{\Ntx \times n_p}$ is a $2$D-DFT (sub)matrix, constructed by the Kronecker product of two \ac{DFT} matrices, $\mbP = \mbP_\text{h} \otimes \mbP_\text{v}$, where each column $\mbp_p$ of $\mbP$, for $p \in \{1,2, \dots, n_p\} $, is normalized such that $\|\mbp_p\|^2=\rho$, since we employ a \ac{URA} at the \ac{BS} \cite{TsZhWa18}. In this work, we consider $n_p\leq \Ntx$.
For what follows, it is convenient to vectorize~\eqref{eq:noisy_obs}:
\begin{equation}
    \mby = \mbA \mbh + \mbn,
\end{equation}
with the definitions \( \mbh = \vect(\mbH) \), \( \mby = \vect(\mbY) \), \( \mbn = \vect(\mbN) \), \( \mbA = \mbP^\tp \otimes \mbI_{\Nrx} \) and $\mbn \sim \mathcal{N}_\C(\mathbf{0}, \mbSigma = \sigma_n^2 \mbI_{n_p \times \Nrx})$.

\subsection{GMM based Channel Estimation}
\label{sec:gmm_chan_est}

A \ac{GMM} is a \ac{PDF} of the form~\cite{bookBi06}
\begin{equation}\label{eq:gmm_of_h}
    f^{(K)}_{\mbh}(\mbh) = \sum_{k=1}^K p(k) \calN_{\C}(\mbh; \mbmu_k, \mbC_k)
\end{equation}
where every summand is one of its \( K \) \textit{components}.
Maximum likelihood estimates of the parameters of a \ac{GMM}, i.e, the means $\mbmu_k$, the covariances $\mbC_k$, and the mixing coefficients $p(k)$, can be computed using a training data set \( \mc{H} \) and an \ac{EM} algorithm, see~\cite{bookBi06}.
\acp{GMM} allow for the evaluation of \textit{responsibilities}~\cite{bookBi06}:
\begin{equation}\label{eq:responsibilities_h}
    p(k \mid \mbh) = \frac{p(k) \calN_{\C}(\mbh; \mbmu_k, \mbC_k)}{\sum_{i=1}^K p(i) \calN_{\C}(\mbh; \mbmu_i, \mbC_i) }.
\end{equation}
These corresponds to the probability that a given \( \mbh \) was drawn from component \( k \).
Interestingly, \acp{GMM} can approximate any continuous \ac{PDF} arbitrary well~\cite{NgNgChMc20}.

With this \ac{GMM} background, we now briefly recap the \ac{GMM} channel estimator from \cite{KoFeTuUt22}.
Given a training data set of channels \(\mathcal{H} = \{ \mbh_m = \vect{(\mbH_m)} \}_{m=1}^{M} \), the \ac{EM} algorithm is used to compute a \( K \)-component \ac{GMM} \( f_{\mbh}^{(K)} \) as an approximation of the true but unknown channel \ac{PDF} \( f_{\mbh} \).

The idea in~\cite{KoFeTuUt22, KoFeTuUt21J} now is to compute the \ac{MSE}-optimal estimator \( \hhat_{\text{GMM}}^{(K)} \) for channels distributed according to \( f_{\mbh}^{(K)} \) and to use it to estimate the channels distributed according to \( f_{\mbh} \).
The motivation for this is that \( \hhat_{\text{GMM}}^{(K)} \) converges pointwise to the \ac{MSE}-optimal estimator for channels distributed according to \( f_{\mbh} \) as \( K \to \infty \)~\cite{KoFeTuUt22, KoFeTuUt21J}.

The estimator \( \hhat_{\text{GMM}}^{(K)} \) can be computed in closed form:
\begin{equation}\label{eq:gmm_estimator_closed_form}
    \hhat_{\text{GMM}}^{(K)}(\mby) = \sum_{k=1}^K p(k \mid \mby) \hhat_{\text{LMMSE},k}(\mby)
\end{equation}
with the responsibilities
\begin{equation}\label{eq:responsibilities}
    p(k \mid \mby) = \frac{p(k) \calN_{\C}(\mby; \mbA \meanhk, \mbA \covhk \mbA^\herm + \mbSigma)}{\sum_{i=1}^K p(i) \calN_{\C}(\mby; \mbA \meanhi, \mbA \covhi \mbA^\herm + \mbSigma) }
\end{equation}
and
\begin{equation}\label{eq:lmmse_formula}
    \hhat_{\text{LMMSE},k}(\mby) =
    \covhk \mbA^\herm (\mbA \covhk \mbA^\herm + \mbSigma)^{-1} (\mby - \mbA \meanhk) + \meanhk.
\end{equation}
The estimator \( \hhat_{\text{GMM}}^{(K)} \) calculates a weighted sum of \( K \) \ac{LMMSE} estimators---one for each component.
The weights \( p(k \mid \mby) \) are the probabilities that the current observation \( \mby \) corresponds to the \(k\)th component.

\subsection{Baseline Channel Estimators}

As a first baseline, we consider a sample covariance matrix based channel estimation approach, where we construct a sample covariance matrix $\mbC_s = \frac{1}{M} \sum_{m=1}^M \mbh_m \mbh_m^\herm$ given the same set of training samples which is used to fit the \ac{GMM} and calculate \ac{LMMSE} channel estimates:
\begin{equation}\label{eq:sample_cov}
    \hhat_{\text{s-cov}} = \mbC_s \mbA^\herm (\mbA \mbC_s \mbA^\herm + \mbSigma)^{-1} \mby.
\end{equation}
Secondly, compressive sensing approaches commonly assume that the channel exhibits a certain structure: $\mbh \approx \mbD \mbt$, where $\mbD = \mbD_{\text{rx}} \otimes (\mbD_{\text{tx,h}} \otimes \mbD_{\text{tx,v}})$ is a \textit{dictionary} with oversampled \ac{DFT} matrices $\mbD_{\text{rx}}$, $\mbD_{\text{tx,h}}$ and $\mbD_{\text{tx,v}}$ (cf., e.g., \cite{AlLeHe15}), because we have a \ac{URA} at the transmitter and a \ac{ULA} at the receiver side.
A compressive sensing algorithm like \ac{OMP}~\cite{Gharavi} can now be used to obtain a sparse vector \( \mbt \), and the estimated channel is then given by
\begin{equation} \label{eq:omp_est}
    \hhat_{\text{OMP}} = \mbD \mbt.
\end{equation}
Since the sparsity order is not known but the algorithm's performance crucially depends on it, we use a genie-aided approach to obtain a bound on the performance of the algorithm. Namely, we use the true channel (perfect \ac{CSI} knowledge) to choose the optimal sparsity order.

\section{Codebook Design}


\subsection{Proposed Codebook Construction and Encoding Scheme}
\label{sec:proposedscheme}

As explained around~\eqref{eq:gmm_estimator_closed_form}, the first step in computing channel estimates via \( \hhat_{\text{GMM}}^{(K)} \) consists of determining how likely it is that the current observation \( \mby \) corresponds to the \( k \)th component of the \ac{GMM} \( f_{\mbh}^{(K)} \), see the responsibility \( p(k\mid \mby) \) in~\eqref{eq:responsibilities}.
The idea of the proposed method now is to compute a codebook transmit covariance matrix \( \mbQ_k \) for every component of the \ac{GMM} and to use the responsiblities \( p(k\mid \mby) \) to determine the feedback index.

In detail, in an offline training phase, we take \( K = 2^B \) as the number of \ac{GMM} components,
use a training data set of channels \( \mathcal{H} = \{\mbh_m \}_{m=1}^{M} \) to fit a \( K \)-component \ac{GMM} \( f_{\mbh}^{(K)} \),
and compute a codebook \( \mc{Q} = \{ \mbQ_k \}_{k=1}^K \) of transmit covariance matrices---one matrix for every \ac{GMM} component.
We explain the codebook construction in another paragraph below.
During the online phase, we bypass explicit channel estimation and directly determine a feedback index using the responsibilities computed via \( \mby \):
\begin{equation} \label{eq:ecsi_index}
    k^\star = \argmax_{k } {p(k \mid \mby)}.
\end{equation}
Thus, we compute the feedback index \( k^\star \) without requiring (estimated) \ac{CSI}.
Note, we thereby also avoid the \( \log_2 \det \) evaluation in~\eqref{eq:codebook_index_selection}.
Further, the knowledge of the codebook at the \ac{MT} is not required. 
The \ac{MT} only requires the \ac{GMM} to compute \eqref{eq:ecsi_index}.

We can think of \( p(k \mid \mby) \) as an approximation of \( p(k \mid \mbh) \) from~\eqref{eq:responsibilities_h}, because of the fixed noise covariance of every component.
That is, since there is a true underlying channel \( \mbh \) leading to the current observation \( \mby = \mbA \mbh + \mbn \), \( p(k \mid \mby) \) can be seen as an approximation of the probability \( p(k \mid \mbh) \) that the channel \( \mbh \) was generated from the \( k \)th \ac{GMM} component.
To both gauge the influence of using \( p(k \mid \mby) \) instead of \( p(k \mid \mbh) \) and to evaluate the codebook itself, it is interesting to look at the performance of feedback information calculated as
\begin{equation} \label{eq:pcsi_index}
    k^\star = \argmax_{k } {p(k \mid \mbh)}.
\end{equation}
Of course, this approach is not practically feasible because the channel \( \mbh \) would have to be known.

\emph{Codebook construction:}
Once the training data set \( \mc{H} = \{\mbh_m \}_{m=1}^{M} \) has been used to fit a \( K \)-component \ac{GMM}, we cluster the training data according to their \ac{GMM} reponsibilities.
That is, we partition \( \mc{H} \) into \( K \) disjoint sets
\begin{equation}\label{eq:lloyd_stage_1}
    \mc{V}_k = \{ \mbh \in \mathcal{H} \mid p(k \mid \mbh) \geq p(j \mid \mbh) \text{ for } k\neq j \}
\end{equation}
for \( k = 1, \dots, K \).
For a channel matrix \( \mbH \) and a covariance matrix \( \mbQ \), let
\begin{equation}
    r(\mbH, \mbQ) = \log_2 \det\left( \mbI + \frac{1}{\sigma_n^2} \mbH \mbQ \mbH^\herm\right)
    \label{speceff}
\end{equation}
be the spectral efficiency.
We now determine the codebook \( \mc{Q} = \{ \mbQ_k \}_{k=1}^K \) by computing every transmit covariance matrix \( \mbQ_k \) such that it maximizes the summed rate in \( \mc{V}_k \):
\begin{align}\label{eq:gmmcb_stage_2}
    &\mbQ_k = \argmax_{\mbQ \succeq \mbzero} \frac{1}{|\mathcal{V}_k|} \sum_{\vect(\mbH)\in\mathcal{V}_k} r(\mbH,\mbQ) \\
    & \text{subject to} \quad \operatorname{trace}(\mbQ) \leq \rho \quad \text{and} \quad
    \operatorname{rank}\mbQ \leq \Nrx. \nonumber
\end{align}
This optimization problem is solved via \ac{PGD}, cf.~\cite{TuKoBaXuUt21, HuScJoUt08}.

In summary, the \ac{GMM} is used twice: Once for codebook construction (done offline) and thereafter to determine a feedback index (done online). For the latter, it is not necessary to estimate the channel and evaluating~\eqref{eq:codebook_index_selection} is avoided.

\subsection{Conventional Codebook Construction Methods}

A standard codebook construction approach makes use of Lloyd's algorithm~\cite{LiBuGr80, LaYoCh04}.
Given a training data set of channels \( \mathcal{H} = \{\mbh_m \}_{m=1}^{M} \), the iterative Lloyd clustering algorithm alternates between two stages until a convergence criterion is met.
We write $ \{ \mbQ_k^{(i)} \}_{k=1}^{\Ncbentries} $ for the codebook in iteration $ i $.
The two stages in iteration $ i $ are:
\begin{enumerate}
    \item Divide the training data set $ \mathcal{H} $ into $ \Ncbentries $ clusters $ \mathcal{V}_k^{(i)} $:
    \begin{equation}\label{eq:lloyd_stage_1_conv}
        \mathcal{V}_k^{(i)} = \{ \mbh \in \mc{H} \mid r(\mbH, \mbQ_k^{(i)}) \geq r(\mbH, \mbQ_j^{(i)}) \text{ for } k\neq j \}.
    \end{equation}
    \item Update the codebook:
    \begin{align}\label{eq:lloyd_stage_2}
        &\mbQ_k^{(i+1)} = \argmax_{\mbQ \succeq \mbzero} \frac{1}{|\mathcal{V}_k^{(i)}|} \sum_{\vect(\mbH)\in\mathcal{V}_k^{(i)}} r(\mbH,\mbQ) \\
        & \text{subject to} \quad \operatorname{trace}(\mbQ) \leq \rho \quad \text{and} \quad
        \operatorname{rank}\mbQ \leq \Nrx. \nonumber
    \end{align}
\end{enumerate}
The optimization problem in stage 2) is again solved via \ac{PGD}.
To initialize the algorithm, stage 1) is replaced with a random partition of \( \mc{H} \) in the first iteration.

\emph{Lau's heuristic:} In order to avoid solving the costly optimization problem in stage 2) of every iteration, the authors of~\cite{LaYoCh04} provide a heuristic for the codebook update: A representative matrix
$\mbS_k^{(i)} = \frac{1}{|\mathcal{V}_k^{(i)}|} \sum_{\vect(\mbH)\in\mathcal{V}_k^{(i)}} \mbH^\herm \mbH$ is calculated for every cluster \( \mc{V}_k^{(i)} \), and then the matrices $\mbS_k^{(i)}$ are decomposed into $\Nrx$ parallel streams and water-filling is employed, yielding the updated codebook entries $\mbQ_k^{(i+1)}$, see~\cite{LaYoCh04}.

Analogously, we can replace the optimization problem in~\eqref{eq:gmmcb_stage_2} with the described heuristic from~\cite{LaYoCh04} to compute a transmit covariance matrix for every \ac{GMM} component.
However, as the simulation results in Section~\ref{sec:sim_results} show, the performance with \ac{PGD} is better, especially in the high \ac{SNR} regime.

\section{Complexity Analysis}

The responsibilities in \eqref{eq:responsibilities_h} or \eqref{eq:responsibilities} are calculated by evaluating Gaussian densities.
A Gaussian density with mean \( \mbmu \in \C^N \) and covariance matrix \( \mbC \in \C^{N\times N} \) can be written as
\begin{equation}\label{eq:gaussian_density}
    \calN_{\C}(\mbh; \mbmu, \mbC) = \frac{\exp(-(\mbh - \mbmu)^\herm \mbC^{-1} (\mbh - \mbmu))}{\pi^N \det(\mbC)}.
\end{equation}
Since the \ac{GMM} covariance matrices and mean vectors do not change between observations, the inverse and the determinant of the densities can be pre-computed once.
Therefore, the online evaluation is dominated by matrix-vector multiplications and has a complexity of \( \calO(N^2) \), where $N=\Ntx \Nrx$ to evaluate \( p(k \mid \mbh) \) in~\eqref{eq:responsibilities_h} (assuming perfect \ac{CSI}), or $N=n_p \Nrx$ to evaluate \( p(k \mid \mby) \) in~\eqref{eq:responsibilities} using the observations $\mby$.

\subsection{Kronecker Approximation for Saving Complexity}

In order for a \ac{MT} to be able to compute feedback indices,
the parameters of the \ac{GMM} \( f_{\mbh}^{(K)} \) need to be offloaded to the \ac{MT} upon entering the \ac{BS}'s coverage area.
As demonstrated in a numerical example below, the number of \ac{GMM} parameters can be quite large.
This is mainly due to the large number of parameters of the \ac{GMM}'s covariance matrices.
As a remedy, we can constrain the \ac{GMM} covariance matrices to a particular form with less parameters.

For spatial correlation scenarios, a well-known assumption is that the scattering in the vicinity of the transmitter and of the receiver are independent of each other, cf.~\cite{KeScPeMoFr02}.
This assumption leads to channel covariance matrices \( \mbC \), which can be decomposed into the Kronecker product of a transmit and receive side spatial covariance matrix: \( \mbC = \mbC_{\text{tx}} \otimes \mbC_{\text{rx}} \).
As in \cite{KoFeTuUt21J}, we use this assumption to construct a \ac{GMM} consisting of Kronecker product covariance matrices \( \covhk = \mbC_{\text{tx},k} \otimes \mbC_{\text{rx},k} \).

The procedure suggested thus far is to fit a single \ac{GMM} using the vectorized channel training data \( \mc{H} = \{ \mbh_m = \vect(\mbH_m) \}_{m=1}^M \) of dimension \( N = \Ntx \Nrx \).
This results in unconstrained \ac{GMM} covariance matrices of dimension \( N \times N \).
To achieve Kronecker product covariance matrices, a two stage procedure is used in~\cite{KoFeTuUt21J}.
First, we fit two independent ``transmit and receive \acp{GMM}'' with respective covariance matrices of dimensions \( \Ntx \times \Ntx \) and \( \Nrx \times \Nrx \).
To this end, all rows of \( \{ \mbH_m \}_{m=1}^M \) are used to fit a \( \Ktx \)-component transmit \ac{GMM}, and all columns of \( \{ \mbH_m \}_{m=1}^M \) are used to fit a \( \Krx \)-component receive \ac{GMM}.
Thereafter, a \( K = \Ktx\Krx \)-component \ac{GMM} with Kronecker covariance matrices of dimension \( N \times N \) is obtained by computing all Kronecker products \( \mbC_{\text{tx},i} \otimes \mbC_{\text{rx},j} \) of the transmit \ac{GMM} covariance matrices \( \mbC_{\text{tx},i} \) and receive \ac{GMM} covariance matrices \( \mbC_{\text{rx},j} \). Please refer to~\cite{KoFeTuUt21J} for more details.
The advantages of the Kronecker \ac{GMM} are a lower offline training complexity, the ability to parallelize the fitting process, and the need for fewer training samples since the Kronecker \ac{GMM} has much fewer parameters.


\emph{Numerical example:}
To illustrate the difference in the number of \ac{GMM} parameters, we plug in the simulation parameters which we consider in Section~\ref{sec:sim_results}.
There, we have, $\Ntx=32$, $\Nrx=16$, $K_\text{tx}=16$ and $K_\text{rx}=4$, which yields $N=\Ntx \Nrx =512$ and $K=\Ktx \Krx=64$.
The normal \ac{GMM} consists of \( K = 64 \) covariance matrices of dimension \( N \times N \) 
which means that it has \( K \frac{N(N+1)}{2} = 8404992 \) covariance parameters (taking symmetries into account). 
By contrast, the Kronecker \ac{GMM} has only \( \Krx \frac{\Nrx(\Nrx+1)}{2} + \Ktx \frac{\Ntx(\Ntx+1)}{2} = 8992 \) covariance parameters. 
Therefore, with the Kronecker \ac{GMM}, the number of parameters which need to be offloaded is drastically reduced.
For this reason, we consider the Kronecker \ac{GMM} in Section~\ref{sec:sim_results}.


\section{Simulation Results}
\label{sec:sim_results}

The \ac{BS} equipped with a \ac{URA} has in total $\Ntx=32$ antenna elements, with $\Ntxv =4$ vertical and $\Ntxh=8$ horizontal elements.
At the \ac{MT} we have a \ac{ULA} with $\Nrx =16$.
We consider $B=6$ feedback bits and thus $ K = 2^6 = 64 $.

We generate datasets with $30 \cdot 10^3$ channels for both the \ac{UL} and \ac{DL} domain of the scenario: $\mathcal{H}^{\text{UL}}$ and $ \mathcal{H}^{\text{DL}} $.
The \ac{UL} channels have a dimension of $32\times16$ and the \ac{DL} channels have a dimension of $16\times32$.
The data samples are normalized such that \( \expec[\|\mbh\|^2] = N = \Ntx \Nrx \) holds for the vectorized channels. We further set $\rho=1$ which
allows us to define the \ac{SNR} as \( \frac{1}{\sigma_n^2} \).
We split the two sets $\mathcal{H}^{\text{UL}}$ and $\mathcal{H}^{\text{DL}}$ into a training set with $M = 20 \cdot 10^3$ samples, and the remaining samples constitute an evaluation set: $\mathcal{H}_{\text{train}}^{\text{UL}}, \mathcal{H}_{\text{eval}}^{\text{UL}}, \mathcal{H}_{\text{train}}^{\text{DL}}, \ \text{and} \ \mathcal{H}_{\text{eval}}^{\text{DL}}.$
However, the \ac{UL} evaluation set $\mathcal{H}_{\text{eval}}^{\text{UL}}$ is not relevant for our considerations and the following transmit strategies are always evaluated on \( \mathcal{H}_{\text{eval}}^{\text{DL}} \), i.e., in the \ac{DL} domain.
When we fit the \ac{GMM} based on \( \mathcal{H}_{\text{train}}^{\text{UL}} \), we transpose all elements of the set to emulate a \ac{DL}.

In the following, we depict the \ac{nSE} as performance measure. The spectral efficiencies achieved with a given transmit covariance matrix are normalized by the spectral efficiency achieved with the optimal transmit covariance matrix which is given by decomposing the channel into $\Nrx$ parallel streams and employing water-filling \cite{Telatar99capacityof}.
The empirical \ac{cCDF} $P(\text{nSE}>s)$ of the normalized spectral efficiency denoted by the variable $s$ (the corresponding random variable is simply denoted by nSE), is used to depict the empirical probability that the \ac{nSE} exceeds a specific value $s$.

We consider the following baseline transmit strategies:

\indent\emph{i}) The curves labeled ``{uni pow cov}'' represent uniform power allocation where the transmit covariance matrix is given by \( \mbQ = \frac{\rho}{\Ntx} \mbI \). 
In this case, no \ac{CSI} knowledge or codebook is used. \\
\indent \emph{ii}) Moreover, ``{uni pow eigsp}'' depicts the transmit strategy where a transmit covariance matrix is calculated by allocating equal power on the eigenvectors of the channel.
That is, the channel is decomposed into $\Nrx$ parallel streams and $\frac{\rho}{\Nrx}$ power is allocated to each stream.
Note, this approach is infeasible because the \ac{BS} would require full knowledge of the \ac{DL} channel (or its eigenvectors).

\begin{figure}[tb]
    \centering
    \begin{tikzpicture}
        \begin{axis}[
            height=\normalplotheight,
            width=\normalplotwidth,
            legend pos= north east,
            legend style={font=\scriptsize, at={(-0.025, 1.4)}, anchor=south west, legend columns=3},
            label style={font=\scriptsize},
            tick label style={font=\scriptsize},
            title style={align=left, font=\scriptsize},
            title={(a) Codebooks @ $\SI{0}{dB}$ SNR, evaluated with perfect CSI},
            xmin=0.3,
            xmax=1,
            ymin=0,
            ymax=1,
            xlabel={normalized spectral efficiency, $s$},
            ylabel={$P(\text{nSE}>s)$},
            grid=both,
        ]
            \addplot[unipowcdf]
                table[x=uniform,y=oneminy_label,col sep=comma] {gmm_feedback/all_se_rel_opt_cdf_gap40_nc64_SNR0dB_ntrain20000.csv};
                \addlegendentry{\pltUniPowCov};

            \addplot[upchansubcdf]
                table[x=unipow_chan_subsp,y=oneminy_label,col sep=comma] {gmm_feedback/all_se_rel_opt_cdf_gap40_nc64_SNR0dB_ntrain20000.csv};
                \addlegendentry{\pltHeigsp};

            \addplot[lloydpgddl]
                table[x=Lloyd_PGD_DL,y=oneminy_label,col sep=comma] {gmm_feedback/all_se_rel_opt_cdf_gap40_nc64_SNR0dB_ntrain20000.csv};
                \addlegendentry{\pltlloydpgddl};
            \addplot[lloydpgdul]
                table[x=Lloyd_PGD_UL,y=oneminy_label,col sep=comma] {gmm_feedback/all_se_rel_opt_cdf_gap40_nc64_SNR0dB_ntrain20000.csv};
                \addlegendentry{\pltlloydpgdul};
            \addplot[lloydlaudl]
                table[x=Lloyd_Lau_DL,y=oneminy_label,col sep=comma] {gmm_feedback/all_se_rel_opt_cdf_gap40_nc64_SNR0dB_ntrain20000.csv};
                \addlegendentry{\pltlloydlaudl};
            \addplot[lloydlauul]
                table[x=Lloyd_Lau_UL,y=oneminy_label,col sep=comma] {gmm_feedback/all_se_rel_opt_cdf_gap40_nc64_SNR0dB_ntrain20000.csv};
                \addlegendentry{\pltlloydlauul};

            \addplot[gmmpgddl]
                table[x=GMM_PGD_DL,y=oneminy_label,col sep=comma] {gmm_feedback/all_se_rel_opt_cdf_gap40_nc64_SNR0dB_ntrain20000.csv};
                \addlegendentry{\pltgmmpgddl};
            \addplot[gmmpgdul]
                table[x=GMM_PGD_UL,y=oneminy_label,col sep=comma] {gmm_feedback/all_se_rel_opt_cdf_gap40_nc64_SNR0dB_ntrain20000.csv};
                \addlegendentry{\pltgmmpgdul};
            \addplot[gmmlaudl]
                table[x=GMM_Lau_DL,y=oneminy_label,col sep=comma] {gmm_feedback/all_se_rel_opt_cdf_gap40_nc64_SNR0dB_ntrain20000.csv};
                \addlegendentry{\pltgmmlaudl};
            \addplot[gmmlauul]
                table[x=GMM_Lau_UL,y=oneminy_label,col sep=comma] {gmm_feedback/all_se_rel_opt_cdf_gap40_nc64_SNR0dB_ntrain20000.csv};
                \addlegendentry{\pltgmmlauul};

        \end{axis}

        \begin{axis}[
            height=\normalplotheight,
            width=\normalplotwidth,
            legend pos= north east,
            legend style={font=\scriptsize, at={(1.1, 1.0)}, anchor=north west},
            label style={font=\scriptsize},
            tick label style={font=\scriptsize},
            title style={align=left, font=\scriptsize},
            title={(b) Codebooks @ $\SI{10}{dB}$ SNR, evaluated with perfect CSI},
            xmin=0.5,
            xmax=1,
            ymin=0,
            ymax=1,
            xlabel={normalized spectral efficiency, $s$},
            ylabel={$P(\text{nSE}>s)$},
            grid=both,
            yshift=-3.5cm,
        ]
            \addplot[unipowcdf]
                table[x=uniform,y=oneminy_label,col sep=comma] {gmm_feedback/all_se_rel_opt_cdf_gap40_nc64_SNR10dB_ntrain20000.csv};
                \addlegendentry{\pltUniPowCov};

            \addplot[upchansubcdf]
                table[x=unipow_chan_subsp,y=oneminy_label,col sep=comma] {gmm_feedback/all_se_rel_opt_cdf_gap40_nc64_SNR10dB_ntrain20000.csv};
                \addlegendentry{\pltHeigsp};

            \addplot[lloydpgddl]
                table[x=Lloyd_PGD_DL,y=oneminy_label,col sep=comma] {gmm_feedback/all_se_rel_opt_cdf_gap40_nc64_SNR10dB_ntrain20000.csv};
                \addlegendentry{\pltlloydpgddl};
            \addplot[lloydpgdul]
                table[x=Lloyd_PGD_UL,y=oneminy_label,col sep=comma] {gmm_feedback/all_se_rel_opt_cdf_gap40_nc64_SNR10dB_ntrain20000.csv};
                \addlegendentry{\pltlloydpgdul};
            \addplot[lloydlaudl]
                table[x=Lloyd_Lau_DL,y=oneminy_label,col sep=comma] {gmm_feedback/all_se_rel_opt_cdf_gap40_nc64_SNR10dB_ntrain20000.csv};
                \addlegendentry{\pltlloydlaudl};
            \addplot[lloydlauul]
                table[x=Lloyd_Lau_UL,y=oneminy_label,col sep=comma] {gmm_feedback/all_se_rel_opt_cdf_gap40_nc64_SNR10dB_ntrain20000.csv};
                \addlegendentry{\pltlloydlauul};

            \addplot[gmmpgddl]
                table[x=GMM_PGD_DL,y=oneminy_label,col sep=comma] {gmm_feedback/all_se_rel_opt_cdf_gap40_nc64_SNR10dB_ntrain20000.csv};
                \addlegendentry{\pltgmmpgddl};
            \addplot[gmmpgdul]
                table[x=GMM_PGD_UL,y=oneminy_label,col sep=comma] {gmm_feedback/all_se_rel_opt_cdf_gap40_nc64_SNR10dB_ntrain20000.csv};
                \addlegendentry{\pltgmmpgdul};
            \addplot[gmmlaudl]
                table[x=GMM_Lau_DL,y=oneminy_label,col sep=comma] {gmm_feedback/all_se_rel_opt_cdf_gap40_nc64_SNR10dB_ntrain20000.csv};
                \addlegendentry{\pltgmmlaudl};
            \addplot[gmmlauul]
                table[x=GMM_Lau_UL,y=oneminy_label,col sep=comma] {gmm_feedback/all_se_rel_opt_cdf_gap40_nc64_SNR10dB_ntrain20000.csv};
                \addlegendentry{\pltgmmlauul};
            \legend{}
        \end{axis}
    \end{tikzpicture}
    \caption{Empirical \acp{cCDF} of the normalized (by the optimal transmit strategy) spectral efficiencies achieved with different codebooks and transmit strategies evaluated with perfect \ac{CSI}.}
    \label{fig:cbatSNR0and10}
\end{figure}
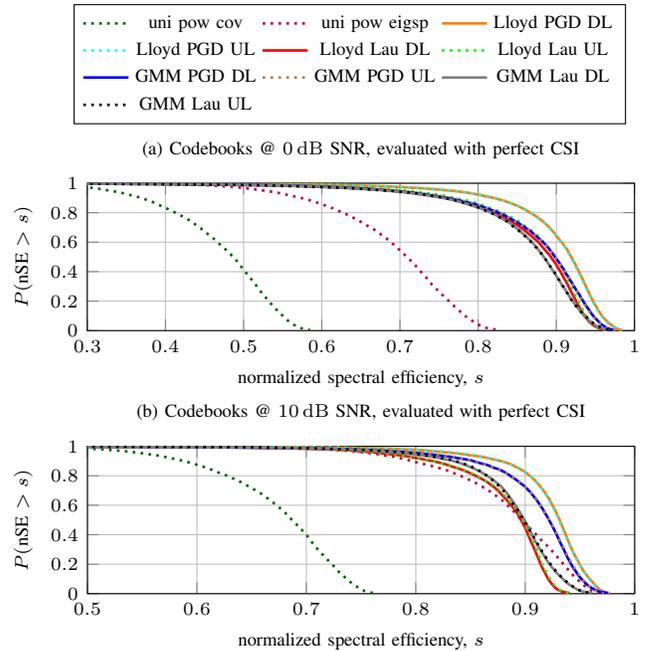

In \Cref{fig:cbatSNR0and10}(a), we set the $\text{SNR}=\SI{0}{dB}$. The conventional codebook construction approaches are denoted by ``{Lloyd PGD UL/DL}'' and ``{Lloyd Lau UL/DL}'', depending on whether \ac{PGD} or Lau's heuristic is used to update the codebook in the second stage of the iterative Lloyd clustering algorithm, and depending on whether $\mathcal{H}_{\text{train}}^{\text{UL}}$ or $\mathcal{H}_{\text{train}}^{\text{DL}}$ is used as training data to construct the codebooks.
With these approaches, the codebook is known to the \ac{BS} and the \ac{MT} and additionally perfect \ac{CSI} is assumed at the \ac{MT}.
Each user then selects the best possible codebook entry by evaluating~\eqref{eq:codebook_index_selection}.
\ac{PGD} seems to be slightly better than Lau's heuristic.
Further, using \ac{DL} or \ac{UL} training data results in approximately the same performance.

The proposed codebook construction and encoding scheme is denoted by ``{GMM PGD UL/DL}'' and ``{GMM Lau UL/DL}'', again depending on whether \ac{PGD} or Lau's heuristic is used to construct the codebook.
We either use $\mathcal{H}_{\text{train}}^{\text{UL}}$ or $\mathcal{H}_{\text{train}}^{\text{DL}}$ as training data to fit the \ac{GMM} and to construct the codebook as described in \ref{sec:proposedscheme}.
With our proposed approach, the knowledge of the codebook at the \ac{MT} is not required.
After offloading the \ac{GMM} to the \ac{MT} and given perfect \ac{CSI} knowledge, the \ac{MT} can then simply determine the feedback index by evaluating \eqref{eq:pcsi_index}.
Again, \ac{PGD} is slightly better than Lau's heuristic, and using \ac{DL} or \ac{UL} training data results in approximately the same performance.
The proposed \ac{GMM} approach performs slightly worse in comparison to the conventional Lloyd clustering approach.
In \Cref{fig:cbatSNR0and10}(b), we set $\text{SNR}=\SI{10}{dB}$, and observe similar results.
Interestingly, ``{GMM Lau UL/DL}'' performs better than ``{Lloyd Lau UL/DL}''.

However, assuming perfect \ac{CSI} at the \ac{MT} is not feasible.
In fact, it is desired to obtain relatively good system performances with estimated \ac{CSI}, where typically only a fraction of the number of transmit antennas $\Ntx$ is used as the number of pilots $n_p$, i.e., when considering systems with reduced pilot overhead.
In the following, we consider the proposed codebook construction and encoding scheme and the conventional Lloyd clustering algorithm exclusively with \ac{PGD} due to its superior performance.
Additionally, we only consider \ac{UL} training data in the remainder.

In \Cref{fig:cbatSNR0P8_SNR15P4}(a), the $\text{SNR}=\SI{0}{dB}$ and we have $n_p=8$.
We depict results for the conventional Lloyd clustering approach, where we first estimate the channel either via \ac{OMP}~\eqref{eq:omp_est}, or the sample covariance approach~\eqref{eq:sample_cov}, or via the \ac{GMM} estimator~\eqref{eq:gmm_estimator_closed_form}, and then select a transmit covariance matrix by evaluating~\eqref{eq:codebook_index_selection} given the estimated channel: ``{{Lloyd PGD}, $\hhat_{\text{OMP}}$}'', ``{{Lloyd PGD}, $\hhat_{\text{s-cov}}$}'', and ``{{Lloyd PGD}, $\hhat_{\text{GMM}}^{(K)}$}''.
As can be seen, estimating the channel via the \ac{GMM} estimator gives the best performance when considering the conventional approach.

By contrast, with our proposed approach denoted by ``{{GMM PGD}, $\mby$}'', where we bypass channel estimation and directly evaluate~\eqref{eq:ecsi_index} for determining a feedback index, we achieve an even better performance as compared to the conventional approach.
With the curves ``{{Lloyd PGD}, $\mbh$}'', and ``{{GMM PGD}, $\mbh$}'' we depict the case of assuming perfect \ac{CSI} knowledge (this is a performance bound).
A similar observation can also be made in \Cref{fig:cbatSNR0P8_SNR15P4}(b), where the $\text{SNR}=\SI{15}{dB}$ and we only have $n_p=4$ pilots.

\begin{figure}[t]
  \centering
  \begin{tikzpicture}
      \begin{axis}[
          height=\smallplotheight,
          width=\smallplotwidth,
          legend pos= north east,
          legend style={font=\scriptsize, at={(-0.25, 1.2)}, anchor=south west, legend columns=3},
          label style={font=\scriptsize},
          tick label style={font=\scriptsize},
          title style={align=left, font=\scriptsize},
          title={(a) SNR $=\SI{0}{dB}$},
          xmin=2,
          xmax=32,
          ymin=0, 
          ymax=1,
          xlabel={$n_p$},
          ylabel={$P(\text{nSE}>0.8)$},
          xtick=data,
          grid=both,
      ]
          \addplot[lloydpgdulperfect,]
              table[x=pilots, y=Lloyd_PGD_UL_perfect_over_pilots, col sep=comma] {gmm_feedback/overpilots/onemincdf_normspeceff0.8_ncomp64_SNR0dB_ntrain20000.csv};
              \addlegendentry{\pltlloydpgdulperfect};
          \addplot[lloydpgdulscov]
                table[x=pilots,y=Lloyd_PGD_UL_scov_over_pilots,col sep=comma] {gmm_feedback/overpilots/onemincdf_normspeceff0.8_ncomp64_SNR0dB_ntrain20000.csv};
                \addlegendentry{\pltlloydpgdulscov};
          \addplot[lloydpgdulomp,]
              table[x=pilots, y=Lloyd_PGD_UL_omp_over_piolts, col sep=comma] {gmm_feedback/overpilots/onemincdf_normspeceff0.8_ncomp64_SNR0dB_ntrain20000.csv};
              \addlegendentry{\pltlloydpgdulomp};
          \addplot[lloydpgdulgmmulall,]
              table[x=pilots, y=Lloyd_PGD_UL_gmmULfull_over_piolts, col sep=comma] {gmm_feedback/overpilots/onemincdf_normspeceff0.8_ncomp64_SNR0dB_ntrain20000.csv};
              \addlegendentry{\pltlloydpgdulgmmulall};
          \addplot[gmmpgdulperfect,]
              table[x=pilots, y=GMM_PGD_UL_perfect_over_pilots, col sep=comma] {gmm_feedback/overpilots/onemincdf_normspeceff0.8_ncomp64_SNR0dB_ntrain20000.csv};
              \addlegendentry{\pltgmmpgdulperfect};
          \addplot[gmmpgdulfromy,]
              table[x=pilots, y=GMM_PGD_UL_from_y_over_pilots, col sep=comma] {gmm_feedback/overpilots/onemincdf_normspeceff0.8_ncomp64_SNR0dB_ntrain20000.csv};
              \addlegendentry{\pltgmmpgdulfromy};
      \end{axis}

      \begin{axis}[
          height=\smallplotheight,
          width=\smallplotwidth,
          legend pos= north east,
          legend style={font=\scriptsize, at={(1.2, 1.0)}, anchor=north west},
          label style={font=\scriptsize},
          tick label style={font=\scriptsize},
          title style={align=left, font=\scriptsize},
          title={(b) SNR $=\SI{5}{dB}$},
          xmin=2,
          xmax=32,
          ymin=0, 
          ymax=1,
          xlabel={$n_p$},
          ylabel={$P(\text{nSE}>0.8)$},
          xtick=data,
          grid=both,
          xshift=4.5cm,
      ]
          \addplot[lloydpgdulperfect,]
              table[x=pilots, y=Lloyd_PGD_UL_perfect_over_pilots, col sep=comma] {gmm_feedback/overpilots/onemincdf_normspeceff0.8_ncomp64_SNR5dB_ntrain20000.csv};
              \addlegendentry{\pltlloydpgdulperfect};
          \addplot[lloydpgdulscov]
                table[x=pilots,y=Lloyd_PGD_UL_scov_over_pilots,col sep=comma] {gmm_feedback/overpilots/onemincdf_normspeceff0.8_ncomp64_SNR5dB_ntrain20000.csv};
                \addlegendentry{\pltlloydpgdulscov};
          \addplot[lloydpgdulomp,]
              table[x=pilots, y=Lloyd_PGD_UL_omp_over_piolts, col sep=comma] {gmm_feedback/overpilots/onemincdf_normspeceff0.8_ncomp64_SNR5dB_ntrain20000.csv};
              \addlegendentry{\pltlloydpgdulomp};
          \addplot[lloydpgdulgmmulall,]
              table[x=pilots, y=Lloyd_PGD_UL_gmmULfull_over_piolts, col sep=comma] {gmm_feedback/overpilots/onemincdf_normspeceff0.8_ncomp64_SNR5dB_ntrain20000.csv};
              \addlegendentry{\pltlloydpgdulgmmulall};
          \addplot[gmmpgdulperfect,]
              table[x=pilots, y=GMM_PGD_UL_perfect_over_pilots, col sep=comma] {gmm_feedback/overpilots/onemincdf_normspeceff0.8_ncomp64_SNR5dB_ntrain20000.csv};
              \addlegendentry{\pltgmmpgdulperfect};
          \addplot[gmmpgdulfromy,]
              table[x=pilots, y=GMM_PGD_UL_from_y_over_pilots, col sep=comma] {gmm_feedback/overpilots/onemincdf_normspeceff0.8_ncomp64_SNR5dB_ntrain20000.csv};
              \addlegendentry{\pltgmmpgdulfromy};
          \legend{}
      \end{axis}
  \end{tikzpicture}
  \caption{The probability that the \ac{nSE} of a certain transmit strategy exceeds $80\%$ of the optimal transmit strategy's spectral efficiency for a varying number of pilots.}
  \label{fig:ccdf_overpilots_32x16_snr0_snr5}
\end{figure}
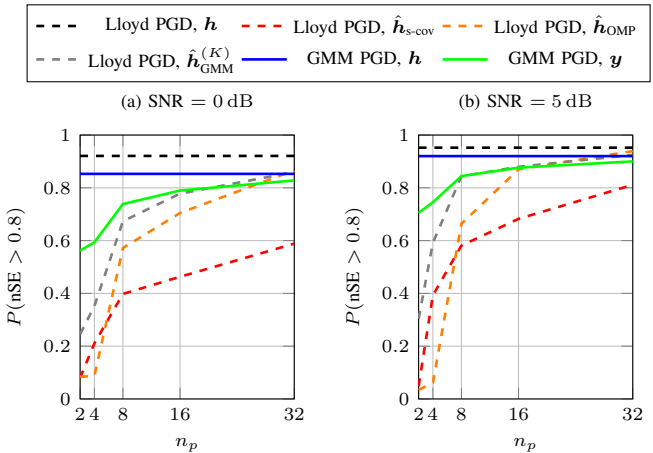

In \Cref{fig:ccdf_overpilots_32x16_snr0_snr5}(a), we set $\text{SNR}=\SI{0}{dB}$ and in \Cref{fig:ccdf_overpilots_32x16_snr0_snr5}(b) we have $\text{SNR}=\SI{5}{dB}$, we fix $s=0.8$ and consider $P(\text{nSE}>0.8)$ for a varying number of pilots $n_p$.
We see, that our proposed approach is especially beneficial in the low number of pilots regime and outperforms the conventional approach, which requires both channel estimation and the evaluation of~\eqref{eq:codebook_index_selection}.

\begin{figure}[tb]
    \centering
    \begin{tikzpicture}
        \begin{axis}[
            height=\normalplotheight,
            width=\normalplotwidth,
            legend pos= north east,
            legend style={font=\scriptsize, at={(-0.125, 1.4)}, anchor=south west, legend columns=3},
            label style={font=\scriptsize},
            tick label style={font=\scriptsize},
            title style={align=left, font=\scriptsize},
            title={(a) Codebooks @ $\SI{0}{dB}$ SNR, evaluated with imperfect CSI @ $n_p=8$},
            xmin=0.3,
            xmax=1,
            ymin=0,
            ymax=1,
            xlabel={normalized spectral efficiency, $s$},
            ylabel={$P(\text{nSE}>s)$},
            grid=both,
        ]
            \addplot[unipowcdf]
                table[x=uniform,y=oneminy_label,col sep=comma] {gmm_feedback/P8/UL_all_se_rel_opt_cdf_gap40_nc64_SNR0dB_ntrain20000.csv};
                \addlegendentry{uni pow cov};

            \addplot[upchansubcdf]
                table[x=unipow_chan_subsp,y=oneminy_label,col sep=comma] {gmm_feedback/P8/UL_all_se_rel_opt_cdf_gap40_nc64_SNR0dB_ntrain20000.csv};
                \addlegendentry{\pltHeigsp};

            \addplot[lloydpgdulperfect]
                table[x=Lloyd_PGD_UL_perfect,y=oneminy_label,col sep=comma] {gmm_feedback/P8/UL_all_se_rel_opt_cdf_gap40_nc64_SNR0dB_ntrain20000.csv};
                \addlegendentry{\pltlloydpgdulperfect};
            \addplot[lloydpgdulscov]
                table[x=Lloyd_PGD_UL_scovUL,y=oneminy_label,col sep=comma] {gmm_feedback/P8/UL_all_se_rel_opt_cdf_gap40_nc64_SNR0dB_ntrain20000.csv};
                \addlegendentry{\pltlloydpgdulscov};
            \addplot[lloydpgdulomp]
                table[x=Lloyd_PGD_UL_omp,y=oneminy_label,col sep=comma] {gmm_feedback/P8/UL_all_se_rel_opt_cdf_gap40_nc64_SNR0dB_ntrain20000.csv};
                \addlegendentry{\pltlloydpgdulomp};
            \addplot[lloydpgdulgmmulall]
                table[x=Lloyd_PGD_UL_gmmULfull,y=oneminy_label,col sep=comma] {gmm_feedback/P8/UL_all_se_rel_opt_cdf_gap40_nc64_SNR0dB_ntrain20000.csv};
                \addlegendentry{\pltlloydpgdulgmmulall};

            \addplot[gmmpgdulperfect]
                table[x=GMM_PGD_UL_perfect,y=oneminy_label,col sep=comma] {gmm_feedback/P8/UL_all_se_rel_opt_cdf_gap40_nc64_SNR0dB_ntrain20000.csv};
                \addlegendentry{\pltgmmpgdulperfect};
            \addplot[gmmpgdulfromy]
                table[x=GMM_PGD_UL_from_y,y=oneminy_label,col sep=comma] {gmm_feedback/P8/UL_all_se_rel_opt_cdf_gap40_nc64_SNR0dB_ntrain20000.csv};
                \addlegendentry{\pltgmmpgdulfromy};

        \end{axis}

        \begin{axis}[
            height=\normalplotheight,
            width=\normalplotwidth,
            legend pos= north east,
            legend style={font=\scriptsize, at={(1.1, 1.0)}, anchor=north west},
            label style={font=\scriptsize},
            tick label style={font=\scriptsize},
            title style={align=left, font=\scriptsize},
            title={(b) Codebooks @ $\SI{15}{dB}$ SNR, evaluated with imperfect CSI @ $n_p=4$},
            xmin=0.5,
            xmax=1,
            ymin=0.0,
            ymax=1,
            xlabel={normalized spectral efficiency, $s$},
            ylabel={$P(\text{nSE}>s)$},
            grid=both,
            yshift=-3.5cm,
        ]
            \addplot[unipowcdf]
                table[x=uniform,y=oneminy_label,col sep=comma] {gmm_feedback/P4/UL_all_se_rel_opt_cdf_gap40_nc64_SNR15dB_ntrain20000.csv};
                \addlegendentry{uni pow cov};

            \addplot[upchansubcdf]
                table[x=unipow_chan_subsp,y=oneminy_label,col sep=comma] {gmm_feedback/P4/UL_all_se_rel_opt_cdf_gap40_nc64_SNR15dB_ntrain20000.csv};
                \addlegendentry{\pltHeigsp};

            \addplot[lloydpgdulperfect]
                table[x=Lloyd_PGD_UL_perfect,y=oneminy_label,col sep=comma] {gmm_feedback/P4/UL_all_se_rel_opt_cdf_gap40_nc64_SNR15dB_ntrain20000.csv};
                \addlegendentry{\pltlloydpgdulperfect};
            \addplot[lloydpgdulomp]
                table[x=Lloyd_PGD_UL_omp,y=oneminy_label,col sep=comma] {gmm_feedback/P4/UL_all_se_rel_opt_cdf_gap40_nc64_SNR15dB_ntrain20000.csv};
                \addlegendentry{\pltlloydpgdulomp};
            \addplot[lloydpgdulscov]
                table[x=Lloyd_PGD_UL_scovUL,y=oneminy_label,col sep=comma] {gmm_feedback/P4/UL_all_se_rel_opt_cdf_gap40_nc64_SNR15dB_ntrain20000.csv};
                \addlegendentry{\pltlloydpgdulscov};
            \addplot[lloydpgdulgmmulall]
                table[x=Lloyd_PGD_UL_gmmULfull,y=oneminy_label,col sep=comma] {gmm_feedback/P4/UL_all_se_rel_opt_cdf_gap40_nc64_SNR15dB_ntrain20000.csv};
                \addlegendentry{\pltlloydpgdulgmmulall};

            \addplot[gmmpgdulperfect]
                table[x=GMM_PGD_UL_perfect,y=oneminy_label,col sep=comma] {gmm_feedback/P4/UL_all_se_rel_opt_cdf_gap40_nc64_SNR15dB_ntrain20000.csv};
                \addlegendentry{\pltgmmpgdulperfect};
            \addplot[gmmpgdulfromy]
                table[x=GMM_PGD_UL_from_y,y=oneminy_label,col sep=comma] {gmm_feedback/P4/UL_all_se_rel_opt_cdf_gap40_nc64_SNR15dB_ntrain20000.csv};
                \addlegendentry{\pltgmmpgdulfromy};
            \legend{}
        \end{axis}
    \end{tikzpicture}
    \caption{Empirical \acp{cCDF} of the normalized (by the optimal transmit strategy) spectral efficiencies achieved with different codebooks and transmit strategies evaluated with imperfect \ac{CSI}.}    
    \label{fig:cbatSNR0P8_SNR15P4}
\end{figure}
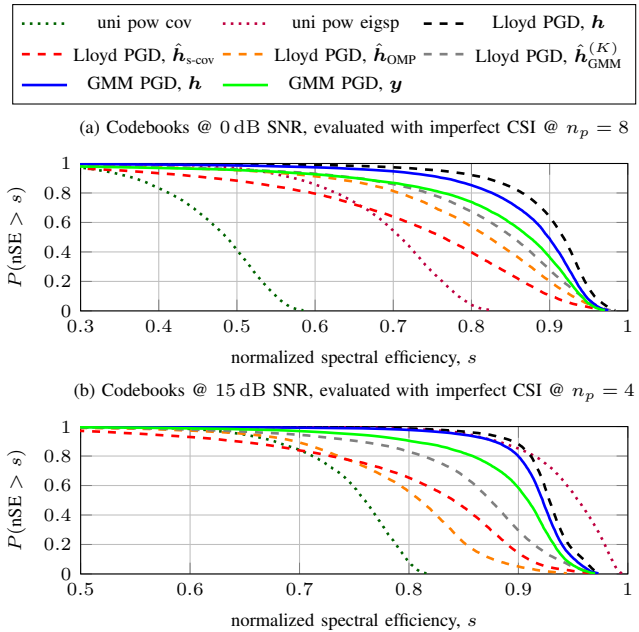

\section{Conclusion and Outlook}

We proposed a codebook construction and feedback encoding scheme which is based on \acp{GMM}. 
The proposed approach involves an offline phase where a \ac{GMM} is fitted and a codebook is constructed at the \ac{BS} using solely \ac{UL} data.
In the online phase, the same \ac{GMM}, which is offloaded to a \ac{MT} upon entering the coverage area of the \ac{BS}, was used for feedback encoding.
Simulation results confirmed the validity of this approach, especially in configurations with a reduced pilot overhead.
In future work, we will investigate the \ac{GMM} based feedback encoding principle for systems with multiple users.

\balance
\bibliographystyle{IEEEtran}
\bibliography{IEEEabrv,biblio}

\end{document}

%% file: miko.tex
\usepackage{mathtools}	
\usepackage{amssymb}
\usepackage{amsthm}

\usepackage[utf8]{inputenc}

\usepackage{ifthen}

\usepackage{microtype}

\usepackage{caption}

\usepackage{bm}

\usepackage[draft,nomargin,marginclue,inline]{fixme}
\makeatletter
\renewcommand*\FXLayoutMarginClue[3]{%
  \marginpar[%
  \raggedleft\@fxuseface{margin}\textcolor{red}{\ignorespaces $ \Rightarrow $}]{%
    \raggedright\@fxuseface{margin}\textcolor{red}{\ignorespaces $ \Leftarrow $}}}
\makeatother	

\usepackage{tumcolor}

\usepackage{pgfplotstable}	

\pgfplotsset{
	discard if/.style 2 args={
        x filter/.append code={
            \edef\tempa{\thisrow{#1}}
            \edef\tempb{#2}
            \ifx\tempa\tempb
                
            \fi
        }
    },
    discard if not/.style 2 args={
        x filter/.append code={
            \edef\tempa{\thisrow{#1}}
            \edef\tempb{#2}
            \ifx\tempa\tempb
            \else
                
            \fi
        }
    }
}

\usepackage{cite}

\usepackage[acronym,shortcuts]{glossaries}
\newacronym{cnn}{CNN}{convolutional neural network}
\newacronym{ula}{ULA}{uniform linear array}

\usepackage{cleveref}

\usepackage{enumitem}

\tikzset{algorithm1/.style={mark options={solid},color=TUMBeamerBlue, line width=\lineWidth, mark=square, dashed}}

\pgfplotscreateplotcyclelist{miko}{%
{color=TUMBeamerRed, dash pattern=on 5pt off 1pt on 1pt off 1pt, line width=1.5pt},%
{color=TUMBeamerOrange, dash pattern=on 5pt off 2pt, line width=1.5pt,
mark=o},%
{color=black, dash pattern=on 5pt off 2pt on 3pt off 2pt, line width=1.5pt},%
{color=TUMDarkerBlue, line width=1.5pt},%
{color=black, dotted, line width=1.5pt},%
{color=TUMBeamerGreen, dotted, line width=1.5pt},%
{color=TUMBeamerYellow, dotted, line width=1.5pt},%
{color=TUMBeamerDarkRed, dotted, line width=1.5pt},%
{color=TUMBeamerLightGreen, dotted, line width=1.5pt},%
{color=TUMIvory, dotted, line width=1.5pt},%
{color=TUMLightBlue, dotted, line width=1.5pt},%
}

\DeclareMathOperator*{\argmax}{arg\,max}

\DeclareMathOperator{\expec}{E}

\DeclareMathOperator{\tr}{tr}

\DeclareMathOperator{\vect}{vec}

\newcommand*{\mc}[1]{\mathcal{#1}}	

\newcommand{\calN}{\mathcal{N}}
\newcommand{\calO}{\mathcal{O}}

\newcommand*{\C}{\mathbb{C}}

\newcommand{\herm}{{\operatorname{H}}}
\newcommand{\tp}{{\operatorname{T}}}

\definecolor{myblue}{RGB}{30, 100, 200}

\newlength{\leftstackrelawd}
\newlength{\leftstackrelbwd}
\def\leftstackrel#1#2{\settowidth{\leftstackrelawd}%
	{${{}^{#1}}$}\settowidth{\leftstackrelbwd}{$#2$}%
	\addtolength{\leftstackrelawd}{-\leftstackrelbwd}%
	\leavevmode\ifthenelse{\lengthtest{\leftstackrelawd>0pt}}%
	{\kern-.5\leftstackrelawd}{}\mathrel{\mathop{#2}\limits^{#1}}}

\newcommand{\mbA}{\bm{A}}

\newcommand{\mbC}{\bm{C}}
\newcommand{\mbD}{\bm{D}}

\newcommand{\mbG}{\bm{G}}
\newcommand{\mbH}{\bm{H}}
\newcommand{\mbI}{\bm{I}}

\newcommand{\mbN}{\bm{N}}

\newcommand{\mbP}{\bm{P}}
\newcommand{\mbQ}{\bm{Q}}

\newcommand{\mbS}{\bm{S}}

\newcommand{\mbY}{\bm{Y}}

\newcommand{\mbh}{\bm{h}}

\newcommand{\mbn}{\bm{n}}

\newcommand{\mbp}{\bm{p}}

\newcommand{\mbs}{\bm{s}}
\newcommand{\mbt}{\bm{t}}

\newcommand{\mbx}{\bm{x}}
\newcommand{\mby}{\bm{y}}

\newcommand{\mbSigma}{{\bm{\Sigma}}}

\newcommand{\mbmu}{{\bm{\mu}}}

\newcommand{\mbzero}{{\bm{0}}}

\newcommand{\hhat}{\hat{\mbh}}

\newcommand{\covhi}{\mbC_i}
\newcommand{\covhk}{\mbC_k}
\newcommand{\meanhi}{\mbmu_i}
\newcommand{\meanhk}{\mbmu_k}

%% file: asilomar22.bbl
\begin{thebibliography}{10}
\providecommand{\url}[1]{#1}
\csname url@samestyle\endcsname
\providecommand{\newblock}{\relax}
\providecommand{\bibinfo}[2]{#2}
\providecommand{\BIBentrySTDinterwordspacing}{\spaceskip=0pt\relax}
\providecommand{\BIBentryALTinterwordstretchfactor}{4}
\providecommand{\BIBentryALTinterwordspacing}{\spaceskip=\fontdimen2\font plus
\BIBentryALTinterwordstretchfactor\fontdimen3\font minus
  \fontdimen4\font\relax}
\providecommand{\BIBforeignlanguage}[2]{{%
\expandafter\ifx\csname l@#1\endcsname\relax
\typeout{** WARNING: IEEEtran.bst: No hyphenation pattern has been}%
\typeout{** loaded for the language `#1'. Using the pattern for}%
\typeout{** the default language instead.}%
\else
\language=\csname l@#1\endcsname
\fi
#2}}
\providecommand{\BIBdecl}{\relax}
\BIBdecl

\bibitem{2019massive}
E.~Bj{\"o}rnson, L.~Sanguinetti, H.~Wymeersch, J.~Hoydis, and T.~L. Marzetta,
  ``Massive {MIMO} is a reality--{W}hat is next? five promising research
  directions for antenna arrays,'' \emph{Digital Signal Processing}, vol.~94,
  pp. 3 -- 20, 2019, special Issue on Source Localization in Massive MIMO.

\bibitem{Love}
D.~J. Love, R.~W. Heath, V.~K. N.~Lau, D.~Gesbert, B.~D. Rao, and M.~Andrews,
  ``An overview of limited feedback in wireless communication systems,''
  \emph{{IEEE} J. Sel. Areas Commun.}, vol.~26, no.~8, pp. 1341--1365, 2008.

\bibitem{utschick2021}
W.~Utschick, V.~Rizzello, M.~Joham, Z.~Ma, and L.~Piazzi, ``Learning the csi
  recovery in fdd systems,'' \emph{{IEEE Trans. on Wireless Commun.}}, 2022.

\bibitem{fesl2021centralized}
B.~Fesl, N.~Turan, M.~Koller, M.~Joham, and W.~Utschick, ``Centralized learning
  of the distributed downlink channel estimators in {FDD} systems using uplink
  data,'' in \emph{WSA 2021; 25th Int. ITG Workshop on Smart Antennas}, 2021,
  pp. 1--6.

\bibitem{TuKoBaXuUt21}
N.~Turan, M.~Koller, S.~Bazzi, W.~Xu, and W.~Utschick, ``Unsupervised learning
  of adaptive codebooks for deep feedback encoding in fdd systems,'' in
  \emph{2021 55th Asilomar Conference on Signals, Systems, and Computers},
  2021, pp. 1464--1469.

\bibitem{TuKoRiFeBaXuUt21}
N.~Turan, M.~Koller, V.~Rizzello, B.~Fesl, S.~Bazzi, W.~Xu, and W.~Utschick,
  ``On distributional invariances between downlink and uplink {MIMO}
  channels,'' in \emph{WSA 2021; 25th Int. ITG Workshop on Smart Antennas},
  2021, pp. 1--6.

\bibitem{NgNgChMc20}
T.~T. Nguyen, H.~D. Nguyen, F.~Chamroukhi, and G.~J. McLachlan, ``Approximation
  by finite mixtures of continuous density functions that vanish at infinity,''
  \emph{Cogent Math. Statist.}, vol.~7, no.~1, p. 1750861, 2020.

\bibitem{LiBuGr80}
Y.~Linde, A.~Buzo, and R.~Gray, ``An algorithm for vector quantizer design,''
  \emph{{IEEE} Trans. Commun.}, vol.~28, no.~1, pp. 84--95, Jan. 1980.

\bibitem{LaYoCh04}
V.~Lau, Y.~Liu, and T.-A. Chen, ``On the design of {MIMO} block-fading channels
  with feedback-link capacity constraint,'' \emph{IEEE Trans. on Commun.},
  vol.~52, no.~1, pp. 62--70, 2004.

\bibitem{Goldsmith}
A.~Goldsmith, \emph{Wireless Communications}.\hskip 1em plus 0.5em minus
  0.4em\relax Cambridge Univ. Press, 2005.

\bibitem{Goldsmith2}
A.~Goldsmith, S.~Jafar, N.~Jindal, and S.~Vishwanath, ``Capacity limits of
  {MIMO} channels,'' \emph{{IEEE} J. Sel. Areas Commun.}, vol.~21, no.~5, pp.
  684--702, 2003.

\bibitem{Telatar99capacityof}
I.~E. Telatar, ``Capacity of multi-antenna gaussian channels,'' \emph{Eur.
  Trans. on Telecommun.}, vol.~10, pp. 585--595, 1999.

\bibitem{QuaDRiGa1}
S.~{Jaeckel}, L.~{Raschkowski}, K.~{Börner}, and L.~{Thiele}, ``Quadriga: A
  3-d multi-cell channel model with time evolution for enabling virtual field
  trials,'' \emph{{IEEE} Trans. Antennas Propag.}, vol.~62, no.~6, pp.
  3242--3256, 2014.

\bibitem{QuaDRiGa2}
S.~{Jaeckel}, L.~{Raschkowski}, K.~{Börner}, L.~{Thiele}, F.~{Burkhardt}, and
  E.~{Eberlein}, ``Quadriga: Quasi deterministic radio channel generator, user
  manual and documentation,'' Fraunhofer Heinrich Hertz Institute, Tech. Rep.,
  v2.2.0, 2019.

\bibitem{TsZhWa18}
Y.~Tsai, L.~Zheng, and X.~Wang, ``Millimeter-wave beamformed full-dimensional
  {MIMO} channel estimation based on atomic norm minimization,'' \emph{IEEE
  Trans. on Commun.}, vol.~66, no.~12, pp. 6150--6163, 2018.

\bibitem{bookBi06}
C.~M. Bishop, \emph{Pattern Recognition and Machine Learning (Information
  Science and Statistics)}.\hskip 1em plus 0.5em minus 0.4em\relax Berlin,
  Heidelberg: Springer-Verlag, 2006.

\bibitem{KoFeTuUt22}
M.~Koller, B.~Fesl, N.~Turan, and W.~Utschick, ``{An Asymptotically Optimal
  Approximation of the Conditional Mean Channel Estimator based on Gaussian
  Mixture Models},'' in \emph{IEEE Int. Conf. on Acoust., Speech and Signal
  Process.}, 2022, pp. 5268--5272.

\bibitem{KoFeTuUt21J}
------, ``{An Asymptotically Optimal Approximation of the Conditional Mean
  Channel Estimator based on Gaussian Mixture Models},'' 2021, arXiv preprint:
  2112.12499.

\bibitem{AlLeHe15}
A.~Alkhateeb, G.~Leus, and R.~W. Heath, ``Compressed sensing based multi-user
  millimeter wave systems: {How} many measurements are needed?'' in \emph{2015
  IEEE Int. Conf. on Acoust., Speech and Signal Process. (ICASSP)}, 2015, pp.
  2909--2913.

\bibitem{Gharavi}
M.~Gharavi-Alkhansari and T.~Huang, ``A fast orthogonal matching pursuit
  algorithm,'' in \emph{Proceedings of the 1998 IEEE Int. Conf. on Acoust.,
  Speech and Signal Process., ICASSP '98 (Cat. No.98CH36181)}, vol.~3, 1998,
  pp. 1389--1392.

\bibitem{HuScJoUt08}
R.~Hunger, D.~A. Schmidt, M.~Joham, and W.~Utschick, ``A general
  covariance-based optimization framework using orthogonal projections,'' in
  \emph{2008 IEEE 9th Workshop on Signal Process. Advances in Wireless
  Commun.}, 2008, pp. 76--80.

\bibitem{KeScPeMoFr02}
J.~Kermoal, L.~Schumacher, K.~Pedersen, P.~Mogensen, and F.~Frederiksen, ``A
  stochastic {MIMO} radio channel model with experimental validation,''
  \emph{{IEEE} J. Sel. Areas Commun.}, vol.~20, no.~6, pp. 1211--1226, 2002.

\end{thebibliography}
